\documentclass[a4paper,11pt]{article}
\usepackage{pos}
\usepackage{epstopdf}
\usepackage{mathrsfs}

\title{The Mini-EUSO telescope on board the International Space Station: Launch and first results}
 \ShortTitle{Mini-EUSO launch and first results }

\author*[a,f,g]{M Casolino}
\author[b,c]{D Barghini}
\author[b,c]{M Battisti}
\author[e]{A Belov}
\author[b,c]{M Bertaina}
\author[b,c]{F Bisconti}
\author[f]{C Blaksley}
\author[h]{K Bolmgren}
\author[p]{F Cafagna}
\author[b,g]{G Cambi\`e}
\author[n]{F Capel}
\author[f]{T Ebisuzaki}
\author[b,c]{F Fenu}
\author[j]{A Franceschi} 
\author[h]{C Fuglesang} 
\author[b,c]{A Golzio}
\author[f]{P Gorodetzki} 
\author[m]{F Kajino} 
\author[f]{H Kasuga}
\author[e]{P Klimov}
\author[r]{V.~Kungel} 
\author[b,c]{M Manfrin}
\author[a]{L Marcelli}
\author[k]{W Marsza{\l}}
\author[b,c]{H Miyamoto}
\author[b,c]{M Mignone} 
\author[j]{T Napolitano}
\author[q]{G Osteria}
\author[l]{E Parizot} 
\author[a,g]{P Picozza} 
\author[o]{L W Piotrowski} 
\author[b]{Z Plebaniak} 
\author[l]{G Pr\'ev\^ot} 
\author[a,g]{E Reali}
\author[j]{M Ricci} 
\author[f]{N Sakaki} 
\author[k]{K Shinozaki} 
\author[k]{J Szabelski} 
\author[f]{Y Takizawa} 
\author[f]{S Wada}
\author[r]{L.~Wiencke}

\affiliation[a]{ INFN, Sezione di Roma Tor Vergata - Roma, Italy}
\affiliation[b]{ INFN, Sezione di Torino - Torino, Italy}
\affiliation[c]{ Dipartimento di Fisica, Universit\'a di Torino, Italy}
\affiliation[d]{ Faculty of Physics, M.V. Lomonosov Moscow State University - Moscow, Russia}
\affiliation[e]{ Skobeltsyn Institute of Nuclear Physics, Lomonosov Moscow State Univ. - Moscow, Russia}
\affiliation[f]{ RIKEN - Wako, Japan}
\affiliation[g]{ Universit\'a degli Studi di Roma Tor Vergata - Dipartimento di Fisica, Roma, Italy}
\affiliation[h]{ KTH Royal Institute of Technology - Stockholm, Sweden}
\affiliation[i]{ Universit\'e de Paris, CNRS, Astroparticule et Cosmologie, F-75006 Paris, France}
\affiliation[j]{ INFN-LNF - Frascati, Italy}
\affiliation[k]{ National Centre for Nuclear Research - Lodz, Poland}
\affiliation[l]{ APC, Univ Paris Diderot, CNRS/IN2P3, CEA/Irfu, Obs de Paris, Sorbonne Paris Cit\'e, France}
\affiliation[m]{ Konan University, Kobe, Japan}
\affiliation[n]{ Technical University of Munich- Munich, Germany} 
\affiliation[o]{ Faculty of Physics, University of Warsaw - Warsaw, Poland}
\affiliation[p]{ INFN, Sezione di Bari - Bari, Italy}
\affiliation[q]{ INFN, Sezione di Napoli - Napoli, Italy}
\affiliation[r]{ Colorado School of Mines, Golden, CO, USA}

\forColl{JEM--EUSO}

\emailAdd{casolino@roma2.infn.it}

\abstract{
Mini-EUSO is a   telescope launched on board the  International Space Station  in 2019 and currently located in the Russian section of the station.  
 Main scientific objectives of the mission are the search for nuclearites and  Strange Quark Matter, the
study of atmospheric phenomena such as Transient Luminous Events, meteors
and meteoroids, the observation of sea bioluminescence and of artificial satellites and man-made space
debris. It is also capable  of observing Extensive Air Showers generated by Ultra-High Energy Cosmic Rays with an energy above 10$^{21}$ eV and detect artificial showers generated with lasers from the ground.
Mini-EUSO can  map the night-time Earth in the UV range (290 - 430 nm), with a spatial resolution of about 6.3 km and a temporal resolution of 2.5 $\mu$s, observing our planet  through a  nadir-facing UV-transparent window in the Russian Zvezda module.
 The instrument, launched on  2019/08/22  from the Baikonur cosmodrome, is based on an optical system employing two Fresnel lenses and  a focal surface composed  of 36 Multi-Anode Photomultiplier tubes, 64 channels each, for a total of 2304 channels with single photon counting sensitivity and an overall  field of view of 44$^{\circ}$.  Mini-EUSO also contains two ancillary cameras to complement  measurements in the near infrared and visible ranges. 
In this paper we describe the detector and present the various phenomena observed in the first year of operation. }

\FullConference{37$^{\rm{th}}$ International Cosmic Ray Conference (ICRC 2021)\\
		July 12th -- 23rd, 2021\\
		Online -- Berlin, Germany}

\begin{document} 
\maketitle

\section{Introduction}
Mini-EUSO (Multiwavelength Imaging New Instrument for the Extreme Universe Space Observatory or 'UV atmosphere' as is known in the Russian Space Program) is a  telescope    operating in the UV range (290 - 430 nm) with a  square field of view of $\simeq $44$^{\circ}$ and a ground resolution of $\simeq 6.3 \times 6.3\:$ km$^2$, depending on the   altitude of the International Space Station (ISS). Mini-EUSO was brought to the  ISS  by the uncrewed Soyuz MS-14, on 2019/08/22. First observations from the  nadir-facing UV transparent window in the Russian Zvezda module  took place  on October $7$. Since then, it has been taking data periodically, with installations occurring every couple of weeks. The instrument is expected to operate for at least three years. For more details on the detector and the data gathered so far see \cite{Mini-EUSO-Astrophys}.

The optical system consists of  two   Fresnel lenses with a  diameter of 25 cm. The focal surface, or Photon Detector Module (PDM), consists of 36 MultiAnode Photomultipliers (MAPMTs) tubes by Hamamatsu, 64 pixels each, capable of single photon detection. Readout is handled by ASICs (Application Specific Integrated Circuit) in frames of $2.5\: \mu s$ (1 Gate Time Unit, GTU). Data are then processed by a Zynq based FPGA board which implements a multi-level triggering, allowing the measurement of triggered UV transients for 128 frames at   time scales of both  $2.5\: \mu$s and $320\: \mu$s (see \cite{matteo} for a description of the trigger system and its performance).  A continuous  acquisition mode  with $\simeq $ 40.96 ms frames is also performed.

\section{Instrument Overview}\label{InstView}
Mini-EUSO has been designed to be installed in the interior of the ISS on the UV-transparent window located in the Zvezda module \ref{fig-scientificobjectives}). The dimensions (37x37x62 cm$^3$) are thus defined by the size of the window and the constraints of the  Soyuz spacecraft. Furthermore, the design accommodates the requirements of safety (no sharp edges, low surface temperature, robustness...) to the crew.   Coupling to the window is done   via a mechanical adapter flange; the only connection to the ISS is via a 28 V power supply and grounding cable. The power consumption of the telescope is $\simeq60\:$ W and the weight is 35 kg, including the  5 kg flange.  
For each observation  session, taking place  about every two weeks and of the duration of about 12 hours,  the instrument is removed from storage and installed on the  UV window. Data are stored   on 512 GB USB Solid State Disks (SSD)  that are inserted in the side of the telescope by the astronauts.  No direct  telecommunication with ground is present, but  samples of data (about 10$\%$, usually corresponding to the beginning and the end of each session) from each session are copied by the crew and transmitted  to ground to verify the correct functioning of the instrument and optimize its working parameters. Conversely, before each session, working parameters, patches in software and hardware are uplinked to the ISS and then copied on the SSD disk to fine-tune the acquisition of the telescope. Pouches with 25 SSDs are   returned    to Earth every 6 months.

The optics  consists of two, 25 cm diameter, Fresnel lenses with a wide field of view (44$^{\circ}$ seen from the PDM).   Poly(methyl methacrylate) - PMMA - is used to manufacture the lenses with a diamond bit machine. In this way it is possible to have a  light (11 mm thickness, 0.87 kg/lens), robust and compact design well suited for space applications.  The effective focal length of the system is 300 mm, with a Point Spread Function (PSF) of 1.2 pixels, of the same dimension as the pixel size of the MAPMTs.

The Mini-EUSO focal surface (PDM) consists of a matrix of 36 Multi-Anode Photomultiplier Tubes (MAPMTs, Hamamatsu Photonics  R11265-M64), arranged in an array of 6$\times $6 elements. Each MAPMT consists of 8$\times$ 8 pixels, resulting in a total of 2304 channels.
The MAPMTs are grouped in Elementary Cells (ECs), each with $2\times2$ units. Each of the nine  ECs  of the PDM shares a common high voltage power supply and a board connecting the dynodes and anodes of the four photomultipliers. The whole system   (250 g each EC, including filters and MAPMTs) is  potted with Arathane and located in the shadow of the photosensors.

\begin{figure}[ht]
\centering
\includegraphics[width=0.6\textwidth]{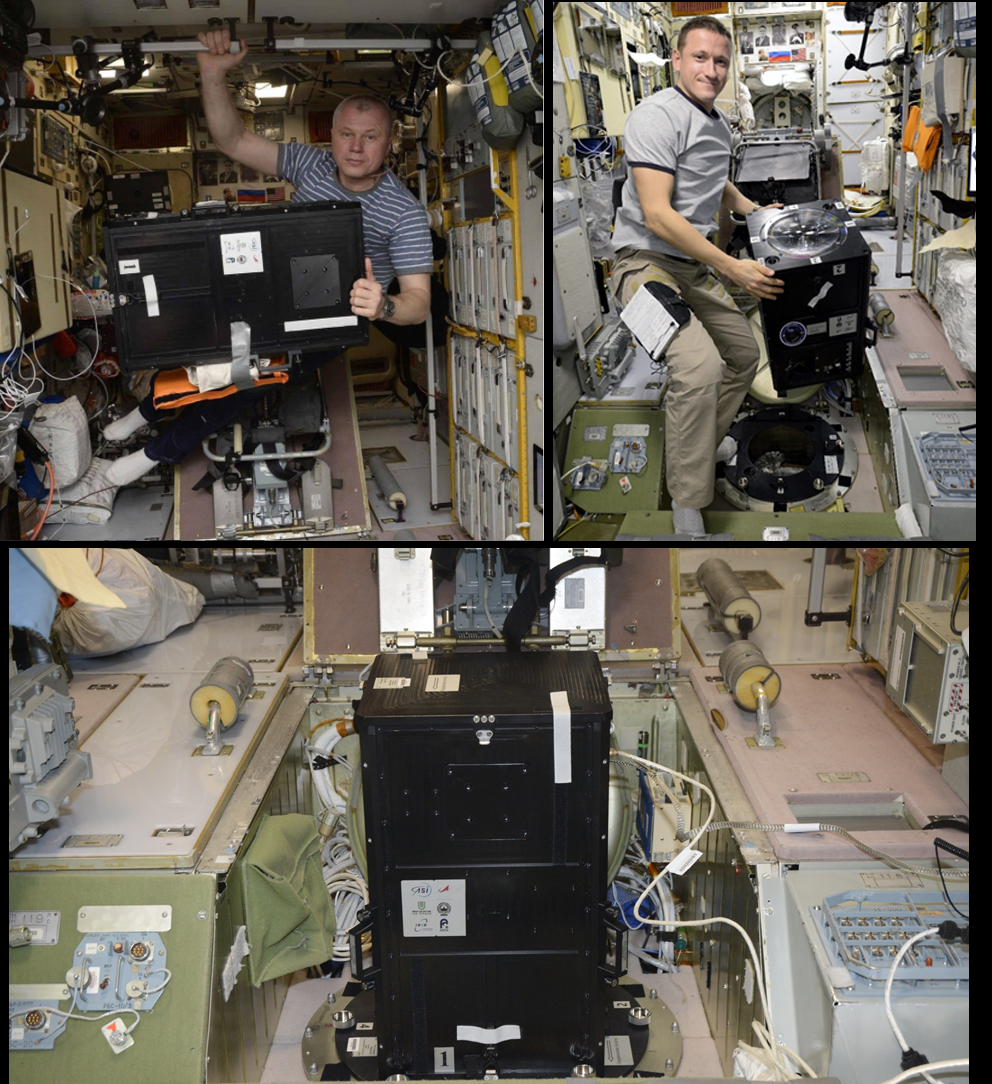}
\caption{Top left and right: Cosmonaut Oleg Novitsky and Sergey Kud-Sverchkov installing Mini-EUSO in the Zvezda module. Bottom: Mini-EUSO installed on the UV transparent window of the  module.   }
\label{fig-photos}       
\end{figure}

After the integration and acceptance tests \cite{giorgio, Mini-EUSO-integration}, first in Rome, subsequently in Moscow and finally in Baikonur cosmodrome, the detector was  integrated in the uncrewed  Soyuz capsule and launched on 2019/08/22. 
 The telescope was first turned on 2019/10/7 (Figure \ref{fig-photos}).   
The first session involved operation in safe mode, with only one EC unit active and the HVPS set  to last dynode voltage mode, corresponding to a sensitivity of about 1$\%$ compared to the normal HVPS mode. Gradually, along the course of the following sessions, the subsequent acquisitions have used the full PDM in normal voltage mode. See \cite{Mini-EUSO-SW} for a description of the software and the acquisition procedure.

\section{Observations}\label{Sci_goals}

\begin{figure}[ht]
\centering
\includegraphics[width=0.6\textwidth]{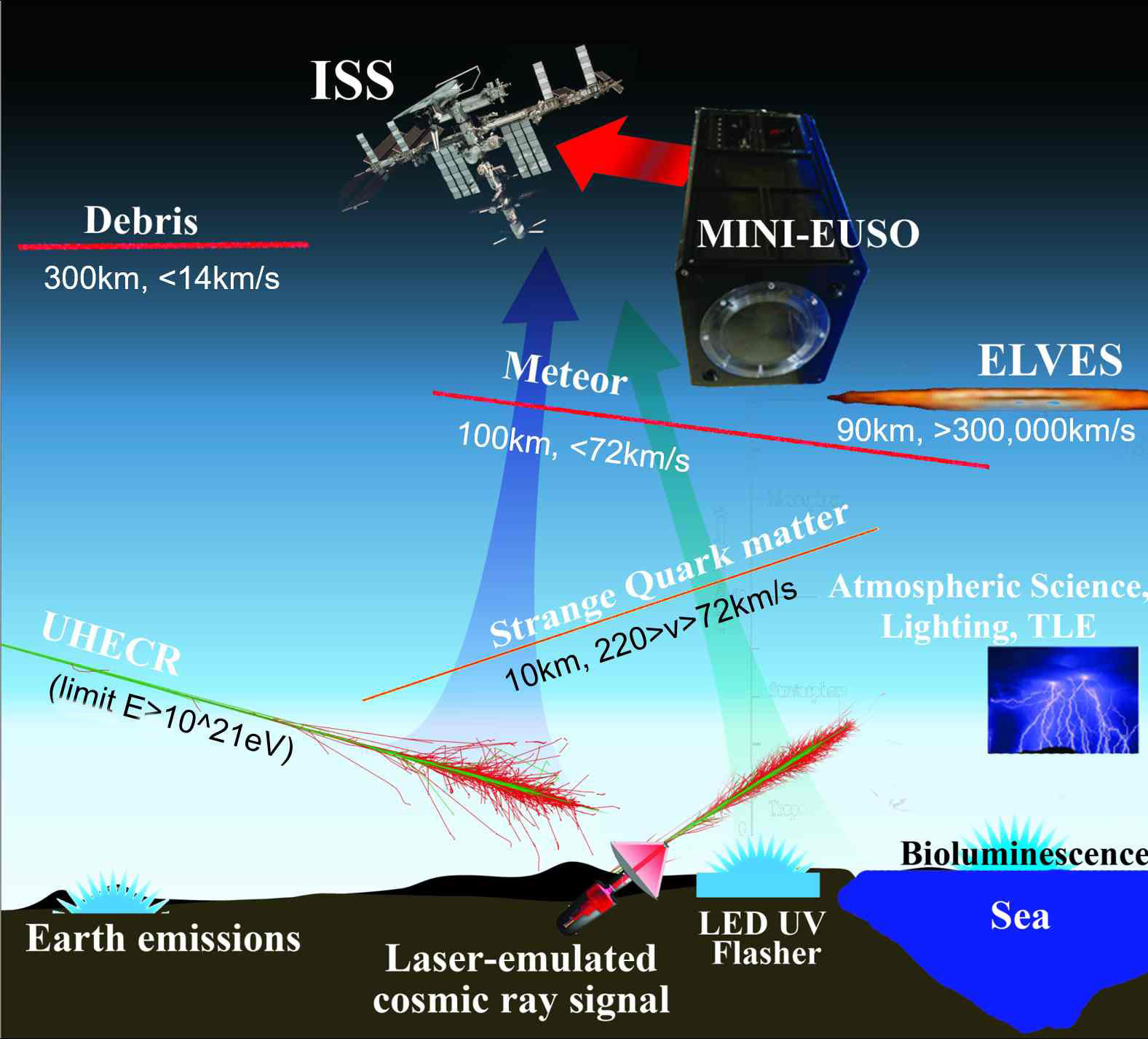}
\caption{Main science goals of  Mini-EUSO. Thanks to its various time acquisitions, ranging from 2.5$\mu s$ to 40ms, the telescope is capable of addressing phenomena with various  duration, from the slow terrestrial emissions (minutes, seconds) to Transient Luminous Events such as  ELVEs ($\simeq 300-400 \mu s$ ).}
\label{fig-scientificobjectives}      
\end{figure}
 
\begin{figure}[ht]
\centering
\includegraphics[width=0.6\textwidth]{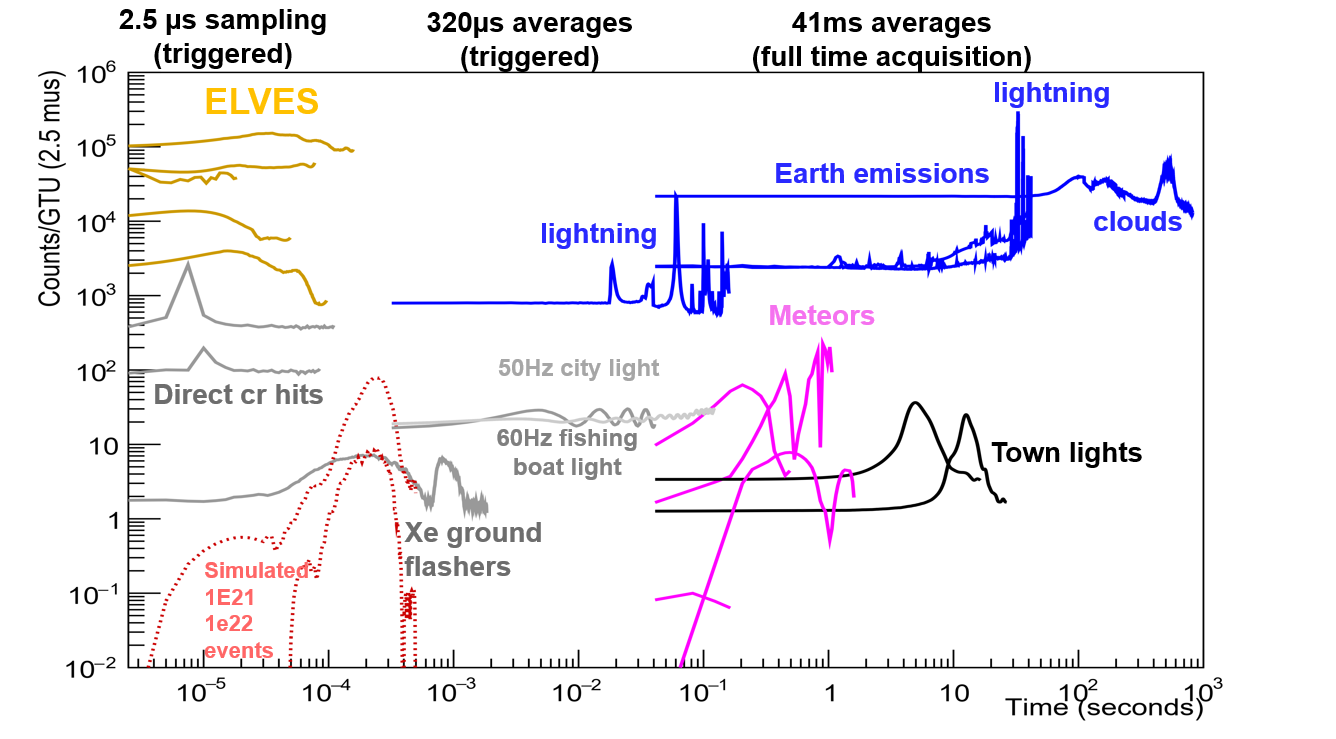}
\caption{Temporal profile of various signals observed by Mini-EUSO. All plots refer to data acquired on the ISS (except the simulations of $10^{21}$ and $10^{22}$ eV UHECRs).}
\label{overview}        
\end{figure}

Mini-EUSO is capable of observations  (Figure \ref{fig-scientificobjectives}) of various phenomena  occurring at various time scales. 
Figure \ref{overview} shows the observed total signal of the focal surface as a function of time for events from the faster $2.5 \mu$s sampling (D1) to the 128 frame averages for D2 (320 $\mu s$) to the 128$\times$128 frame average for D3 (40.96 ms).  In the longer time frames, the gradual increase is due to the passage over a clouded area, whereas the sharp spikes are due to lightning. Large lightning triggers the safety  system of the detector, resulting in the temporary deactivation of the HVPS of the EC unit which would be overexposed by lightning.

\begin{figure}[ht]
\centering
\includegraphics[width=0.6\textwidth]{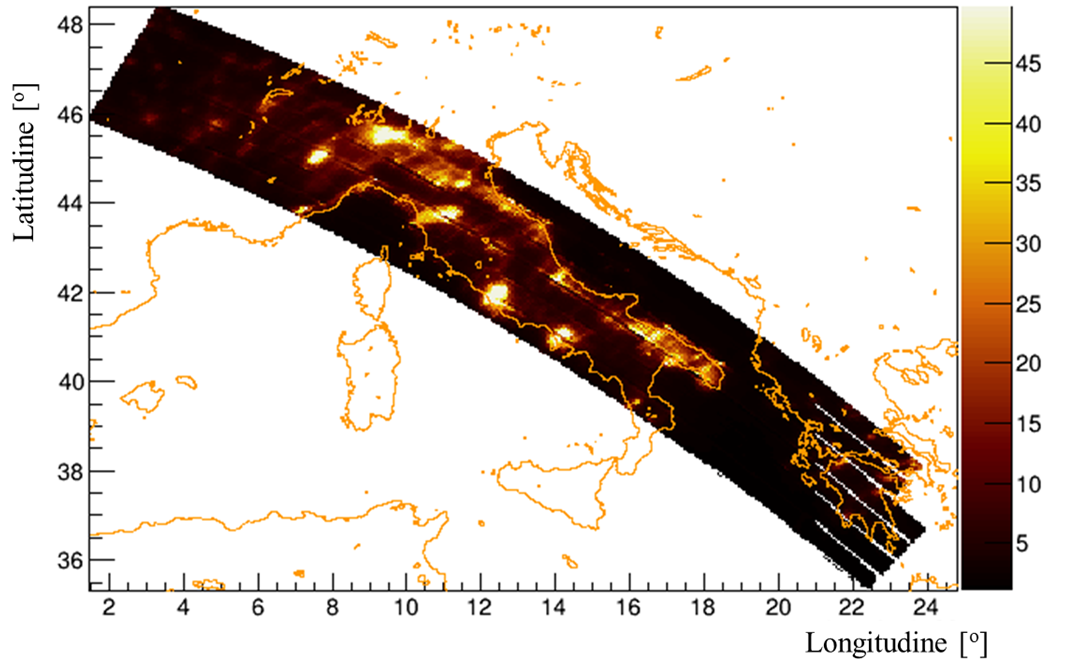}
\caption{Observation of night time emissions from Italy. The light from the various lights can be distinguished, as well as the different emissions from ground and sea. }
\label{fig-nighttime}      
\end{figure}

\begin{figure}[ht]
\centering
\includegraphics[width=0.65\textwidth]{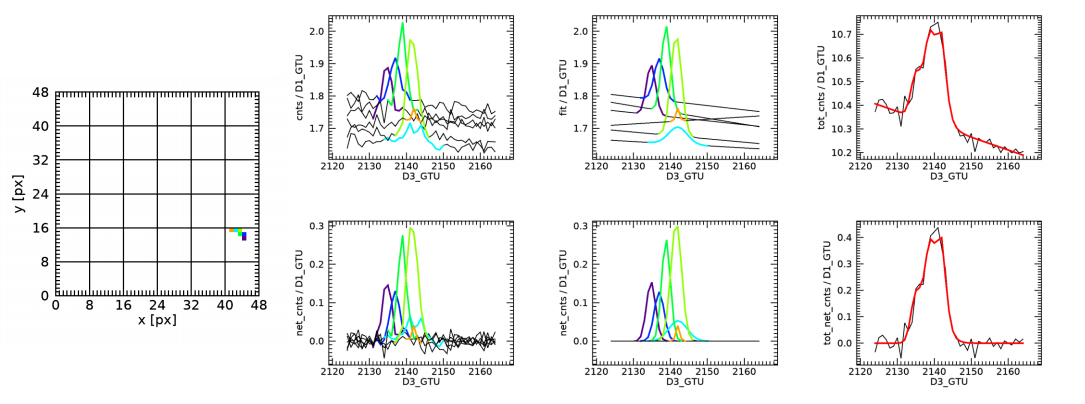}
\caption{Left: A meteor track as it develops in the field of view of Mini-EUSO. The X and Y axis represent the PDM pixels. Right top: the meteor track counts of each pixel, the resulting light curve and, Right bottom: the same curves with background subtracted \cite{dario}.}
\label{fig-Meteor}       
\end{figure}

\begin{figure}[ht]
\centering
\includegraphics[width=0.60\textwidth]{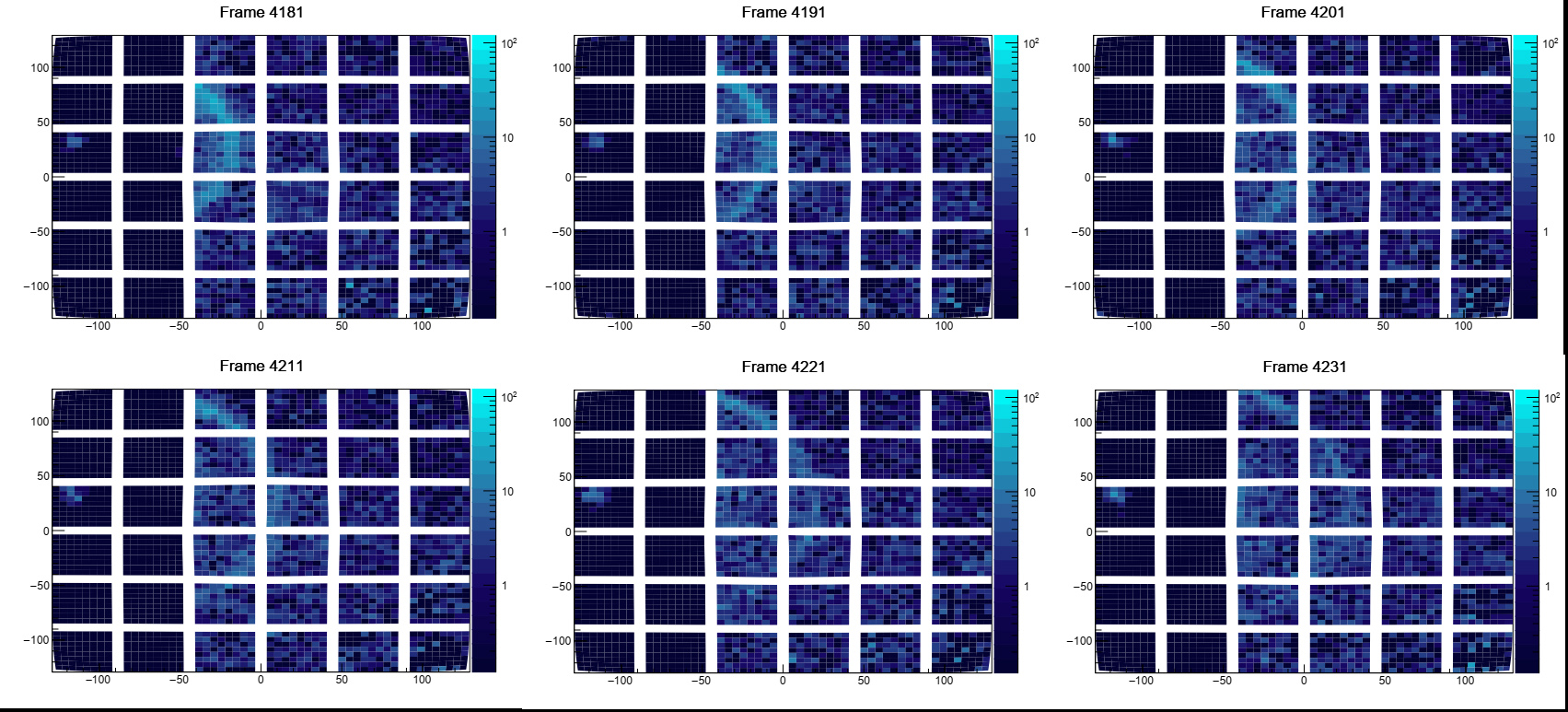}
\caption{A sample of  frames of an ELVE being observed in the focal surface. The left column of three  EC units is temporarily working at a  1/1000 sensitivity due to a previous bright light that triggered the safety mechanism. This allows to see the lightning in the centre left of the focal surface, which generates the expanding elve ring in the centre of the FS. Pictures are 10 frames apart, thus $   2.5\times 10 = 25 \mu s$ apart. }
\label{fig-ELVE}        
\end{figure}

The main observations possible by Mini-EUSO are: 
\begin{enumerate}
\item{Night UV emissions from the Earth.} Mini-EUSO can map the Earth in the near  UV range   with   spatial and temporal resolution  of $\sim 6.3 \times 6.3 $ km$^2$ and 2.5 $\mu$s respectively, measuring variations of the UV emissions.   See Figure \ref{fig-nighttime} for an example of the mapping capabilities of the instrument \cite{alessio,kenji}. Several campaigns with ground LED flashers have been carried out \cite{matteo} or will be carried out with  LED and lasers shooting in the field of view (f.o.v.) of Mini-EUSO.
 
\item{Space debris.} Attempts will be made to track space debris to investigate the possibility of using laser ablation for their removal. The maximum detection distance of Mini-EUSO is about 100 km for debris size of 0.1 m. This observation is restricted to the local twilight period of the orbit, about 5 min every 90 min  \citep{Ebisuzaki2015102ActaAstronautica}.

\item{Meteors} are relatively slow ($v\leq 72$ km/s) and long-lasting  (a few seconds) events which  illuminate in sequence several light sensitive pixels of the Mini-EUSO focal surface (See Figure \ref{fig-Meteor} for the reconstruction of a typical meteor event).   So far we have observed more than 5000 meteor events. The maximum observable magnitude is between 5 and 6 depending on background conditions  \cite{2017P&SS..143..245AMeteorstudiesintheframeworkoftheJEMEUSOprogram,2015ExA....40..253A}.   The frequency-intensity distribution of the observation of meteors will allow to make an inventory of the population of near-Earth objects from space with a large field of view and the advantage of not being covered by clouds. 
Events coming from interstellar meteors (with a kinetic energy above 72 km/s) and Strange Quark Matter (SQM) candidates (appearing as long-lasting, constant-luminosity events) can also be observed \cite{lech}.

\item{Transient Luminous Events, in particular ELVES} are observed as large ring-like upper atmospheric emissions that appear to be expanding at superluminal speed\cite{laura}. In Figure \ref{fig-ELVE} are shown the pictures of one ELVE (observed on 2020/05/26) entering the field of view.
  Furthermore, it will be possible to make joint observations with other detectors on board the ISS such as Altea-Lidal  and ASIM.  

\item{UHECR.} Mini-EUSO can measure fluorescence and Cherenkov light emitted by UHECR initiated showers.   The  diameter of the lens system - constrained by the size of the ISS window - places the threshold energy for UHECR detection  around  10$^{21}$ eV.  We estimate an yearly  exposure  - after commissioning and   selecting observations close to new moon and in regions of low background - to be of the order of  1000 km$^2$ sr yr.  The absence of events with energy above $3\cdot 10^{20} $ eV   obtained with ground detectors   make it unlikely for Mini-EUSO to observe any event at these energies\cite{francesco}. We note that Mini-EUSO exposure is  of the same order of magnitude of the fluorescence detector of Telescope Array, so it can contribute to search for exotic events that would not give a signal in the surface detectors.

\end{enumerate}

\section{Conclusions}\label{Conclusions}
  Initial analysis of the Mini-Euso data received in the first year of operations confirms the correct functioning of the instrument. We have observed events in all the operational time frames, from the fast ELVEs, to meteors (40.96 ms readout), lighting and terrestrial emissions.  
 
\acknowledgments
This work was partially supported by Basic Science Interdisciplinary Research Projects of 
RIKEN and JSPS KAKENHI Grant (22340063, 23340081, and 24244042), by  the Italian Ministry of Foreign Affairs	and International Cooperation, by the Italian Space Agency through the ASI INFN agreements n. 2017-8-H.0 and n. 2021-8-HH.0, by NASA award 11-APRA-0058, 16-APROBES16-0023, 17-APRA17-0066, NNX17AJ82G, NNX13AH54G, 80NSSC18K0246, 80NSSC18K0473, 80NSSC19K0626, and 80NSSC18K0464 in the USA,   by the French space agency CNES, by the Deutsches Zentrum f\"ur Luft- und Raumfahrt, the Helmholtz Alliance for Astroparticle Physics funded by the Initiative and Networking Fund of the Helmholtz Association (Germany), by Slovak Academy of Sciences MVTS JEM-EUSO, by National Science Centre in Poland grants 2017/27/B/ST9/02162 and 2020/37/B/ST9/01821, by Deutsche Forschungsgemeinschaft (DFG, German Research Foundation) under Germany's Excellence Strategy - EXC-2094-390783311, by Mexican funding agencies PAPIIT-UNAM, CONACyT and the Mexican Space Agency (AEM), as well as VEGA grant agency project 2/0132/17, and by by State Space Corporation ROSCOSMOS and the Interdisciplinary Scientific and Educational School of Moscow University "Fundamental and Applied Space Research".

\newpage
{\Large\bf The JEM-EUSO Collaboration\\}
{\scriptsize (author-list as of July 1st, 2021)} \hspace{0.6cm}
\vspace*{0.5cm}

\begin{sloppypar}
{\small \noindent 
G.~Abdellaoui$^{ah}$, 
S.~Abe$^{fq}$, 
J.H.~Adams Jr.$^{pd}$, 
D.~Allard$^{cb}$, 
G.~Alonso$^{md}$, 
L.~Anchordoqui$^{pe}$,
A.~Anzalone$^{eh,ed}$, 
E.~Arnone$^{ek,el}$,
K.~Asano$^{fe}$,
R.~Attallah$^{ac}$
% ...
}
\end{sloppypar}
\vspace*{.3cm}

{ \footnotesize
\noindent
% ...
$^{ac}$ LPR at Dept. of Physics, Faculty of Sciences, University Badji Mokhtar, Annaba, Algeria \\
% ...
$^{ah}$ Telecom Lab., Faculty of Technology, University Abou Bekr Belkaid, Tlemcen, Algeria \\
% ...
$^{cb}$ Universit\'e Paris, CNRS, AstroParticule et Cosmologie, F-75013 Paris, France\\
% ...
$^{ed}$ Istituto Nazionale di Fisica Nucleare - Sezione di Catania, Italy\\
% ...
$^{eh}$ INAF - Istituto di Astrofisica Spaziale e Fisica Cosmica di Palermo, Italy\\
% ...
$^{ek}$ Istituto Nazionale di Fisica Nucleare - Sezione di Torino, Italy\\
$^{el}$ Dipartimento di Fisica, Universita' di Torino, Italy\\
% ...
$^{fe}$ Institute for Cosmic Ray Research, University of Tokyo, Kashiwa, Japan\\ 
% ...
$^{fq}$ Nihon University Chiyoda, Tokyo, Japan\\ 
% ...
$^{md}$ Universidad Polit\'ecnia de madrid (UPM), Madrid, Spain\\
% ...
$^{pd}$ University of Alabama in Huntsville, Huntsville, AL; USA\\
$^{pe}$ Lehman College, City University of New York (CUNY), NY, USA\\
% ...
}

%
%
%


\begin{thebibliography}{99}
\bibitem{Mini-EUSO-Astrophys} S. Bacholle et al, ApJ Supp.,  253, 2, 36, 17, 2020.
 
\bibitem{giorgio}  G. Cambiè et al (JEM-EUSO Coll.), Integration and qualification of the Mini-EUSO telescope on board the ISS, PoS(ICRC2021) 1001.


\bibitem{Mini-EUSO-integration} Belov, A., Cambi\`e, G., Casolino, M., et al. 2020,
Aerotecnica Missili e Spazio, 99, 93, 10.1007/s42496-020-00047-1.
\bibitem{Mini-EUSO-SW} F. Capel et al, JATIS, 5(4), 044009 (2019).
\bibitem{matteo}  M. Battisti et al (JEM-EUSO Coll.), Overview of the Mini-EUSO $\mu$ s trigger logic performance, PoS(ICRC2021) 411. 
\bibitem{alessio} A. Golzio et al (JEM-EUSO Coll.), A study on UV emission from clouds with Mini-EUSO, PoS(ICRC2021) 417.
\bibitem{kenji}  K. Shinozaki et al, PoS(ICRC2021) 1165.



\bibitem{Ebisuzaki2015102ActaAstronautica} Ebisuzaki, T., {et~al.}  Acta Astronautica, 112,   102, 2016 \bibitem{2017P&SS..143..245AMeteorstudiesintheframeworkoftheJEMEUSOprogram} Abdellaoui, G., et al., {2017}, {Planetary and Space Science}, {143}, 245,  {{10.1016/j.pss.2016.12.001}}
\bibitem{2015ExA....40..253A} Adams, J, et al, Experimental Astronomy, 40, 253,
  {10.1007/s10686-014-9375-4}, 2015.
  \bibitem{lech}  L. Piotrowski et al (JEM-EUSO Coll.), Towards observations of nuclearites with Mini-EUSO, PoS(ICRC2021) 1181.
  \bibitem{laura}  L. Marcelli et al (JEM-EUSO Coll.), Observations of Transient Luminous Events with the Mini-EUSO telescope on board the ISS, PoS(ICRC2021) 971.
  \bibitem{francesco}  F. Fenu et al (JEM-EUSO Coll.), Simulation studies for the Mini-EUSO detector, PoS(ICRC2021) 757.
\bibitem{dario} D. Barghini,  14th Europlanet Science Congress 2020,   EPSC2020-800
  September 2020  2020EPSC...14..800B
\end{thebibliography}
\end{document}